\RequirePackage{fix-cm}
\documentclass[twocolumn]{svjour3}          
\smartqed  
\usepackage{graphicx}
\usepackage{ amsmath,amssymb}
\usepackage[section]{placeins}  
\usepackage{dcolumn}
\usepackage{bm}

\usepackage[utf8]{inputenc}
\usepackage[T1]{fontenc}
\usepackage{mathptmx}
\usepackage{subcaption}
\usepackage{booktabs}
\usepackage{tabularx}
\usepackage{makecell}
\usepackage{stfloats}
\usepackage{multicol}
\usepackage{natbib}

\journalname{Experiments in Fluids}
\begin{document}

\title{Measurements of Sub-Surface Velocity Fields in Quasi-2D Faraday Flow
}


\author{Raffaele Colombi  \and
       Michael Schl\"uter  \and
       Alexandra von Kameke
}


\institute{Alexandra von Kameke \at
              Institute of Multiphase Flows\\
              Hamburg University of Technology\\
              Hamburg, Germany  \\
              Tel.: +49 40 42878-4652\\
              \email{alexandra.vonkameke@tuhh.de}\\
              ORCID: 0000-0002-1913-774X           
}

\date{Received: date / Accepted: date}

\maketitle

\begin{abstract}
Faraday waves are capillary ripples that form on the surface of a fluid being subject to vertical shaking. Although it is well known that the form and shape of the waves pattern depend on driving amplitude and frequency,  only recent studies discovered the existence of a horizontal velocity field at the surface, called Faraday flow, which exhibits attributes of two-dimensional turbulence.  
However, despite the increasing attention towards the inverse energy flux in the Faraday flow and other not strictly two-dimensional systems, very little is known about the velocity fields developing beneath the fluid surface. In this study planar velocity fields are measured by means of particle image velocimetry (PIV) with high spatial and temporal resolution on the water surface and below it. A sudden drop in velocity is observed immediately below the water surface, such that at \linebreak 5 mm below the water surface the mean absolute velocities are already about 6.5 times smaller than the surface velocity. Additionally, the flow structures below the surface are found to comprise much larger spatial scales than those on the surface.  These large structures are also found to be slow and temporarily persistent, as proven by analysing the autocorrelation of the velocity fields in time. 
\end{abstract}
\pagebreak
\section{Introduction}
\label{sec:intro}

Faraday waves are capillary ripples that form on the surface of a fluid being subject to vertical agitation. The resulting waves are known to form patterns that vary depending on driving amplitude and frequency \citep{faraday1831xvii}. Because of the strong influence of boundary conditions, Faraday waves are subject to studies for a large variety of applications, ranging from bio-medicine to material sciences (e.g. controlled pattern formation, walking and orbiting of droplets) \citep{saylor2005simulation, couder2005dynamical}.
In capillary ripples, a complex and random transport of floating particles has been accounted to non-linear interactions at the surface of the Faraday wavefield, such as imperfections and traveling waves \citep{saylor2005simulation, ramshankar1990transport}. However, only recent studies \citep{von2011double,von2013measurement, francois2013inverse}, proved the existence of a horizontal velocity field at the surface, called Faraday flow, which was shown to exhibit attributes of two-\linebreak dimensional (2D) turbulence. Recently, the Faraday flow has also been used to control the dispersion of floaters with distinct geometries \citep{francois2018rectification, xia2019tunable, yang2019passive, yang2019diffusion}, opening up a new field of applications and a new study ground for fluid-structure interaction.\\
One of the main features of Faraday flows is the presence of an inverse energy cascade. For 3D isotropic turbulence, energy is injected in the flow at large scales, transported to smaller scales through the vortex stretching mechanism and finally dissipated through viscous effects. However, numerical and experimental results confirmed the presence of a dual energy cascade in case of  2D-turbulence ~(\citep{von2011double, farazmand2011controlling, boffetta2012two, francois2013inverse, liao2013spatial} and references therein), as theoretically predicted in \citep{kraichnan1971inertial}. Energy is introduced at intermediate forcing scales and transferred upwards to larger scales, resulting in a net inverse energy flux. Under particular conditions, this phenomenon can even lead to energy condensation, by which large and ordered flow structures emerge from the the seemingly disordered motion at small scales \citep{xia2009spectrally,musacchio2019condensate, shats2014turbulence}. As forseen by theory, for wavelengths smaller than the forcing scale, an enstrophy cascade transfers enstrophy to the smaller wavelengths.  
Despite the increasing attention towards the well-validated inverse energy flux in the Faraday flow and other not strictly two-dimensional systems \citep{biferale2017two}, very little is known about the flow structures developing beneath the fluid surface in these flows. Recent studies have been conducted in order to assess the three-dimensionality effects in electromagnetically-driven quasi-2D flows \citep{kelley2011onset, martell2019comparing}, and in flows in parametrically-excited waves \citep{francois2014three, xia2017two}, with particular focus on thin-layer flows or in the shallow layers below the waves surface. Additionally, it was recently revealed in \citep{francois2020nonequilibrium} that the mean power extracted from  a rotor placed in a strongly turbulent Faraday flow is highly dependent on the rotor thickness, strongly  influencing the energy of the angular velocity fluctuations. \\
In order to further characterize the complexity of flow-\linebreak structure and wave-fluid interactions it is thus of paramount importance to understand what types of structures are developing beneath the surface, not only in the thin layer where quasi-2D effects prevail, but also at further depths, where structures are expected to be more influenced by bottom friction. Therefore, this study aims at shedding light on the flow characteristics of the Faraday experiment, with particular focus on the velocities below the fluid surface and the resulting flow structures. The velocity fields are measured by means of planar PIV with high spatial and temporal resolution at multiple horizontal planes at different heights for two distinct forcing amplitudes. 

\section{Materials and Methods}
\label{sec:materials_and_methods}

\subsection{\label{sec:container}Water Container and Shaker Set-Up}
\begin{figure}[h!]
		 \centering
	 	 \begin{subfigure}[b]{0.47\textwidth}
       		\centering
	 		\includegraphics[width=\textwidth]{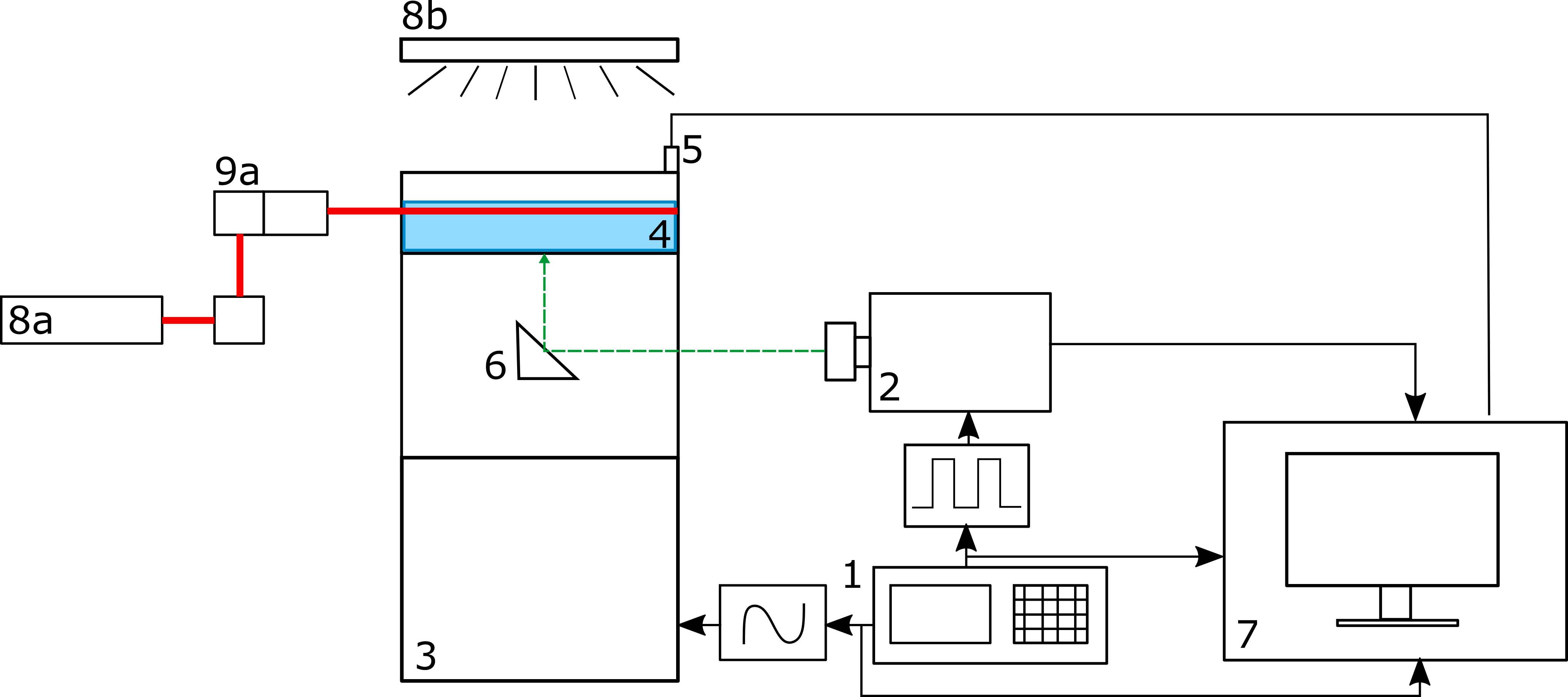}
			\caption{}
		\end{subfigure}
		
		\phantom{a}
		\begin{subfigure}[b]{0.47\textwidth}
			\centering
	 		 \includegraphics[width=0.4\textwidth]{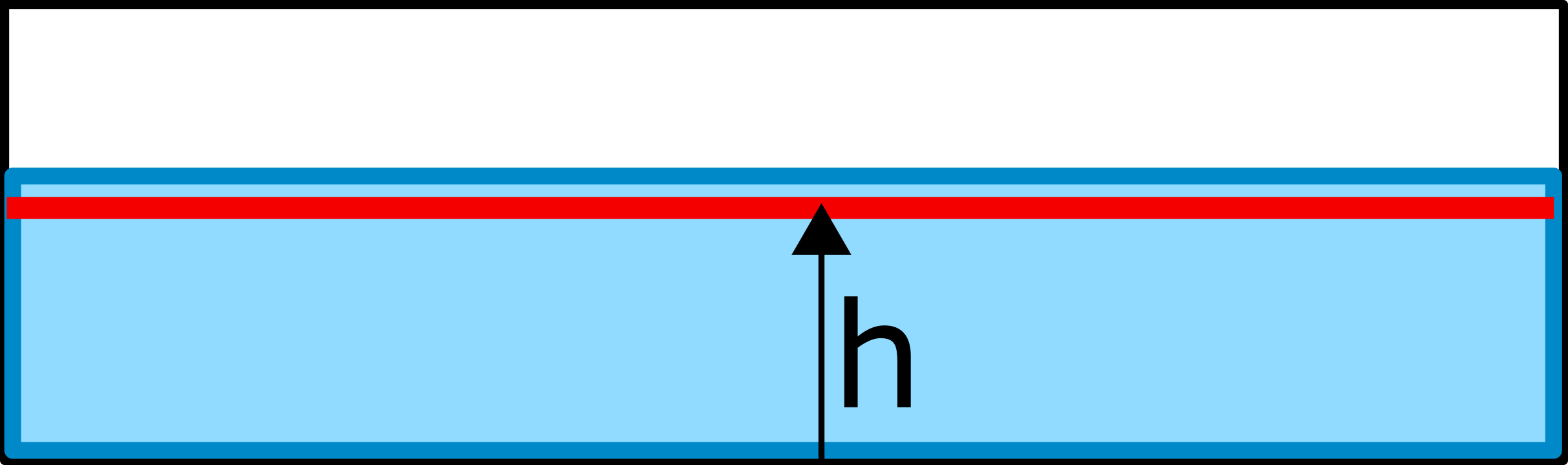}
			\caption{}
			\end{subfigure}

\setlength{\abovecaptionskip}{0pt}			
	 \caption{a) Schematic representation of the experimental set-up. A function generator (1) triggers the high-speed camera (2) and drives the shaker (3). The acceleration of the water container (4) is measured with accelerometers (5). A prism-mirror (6) is used to deflect the camera field of view. All the signals are monitored with a digitizer (7), whereas data is saved on a laboratory computer. Numbers (8a) and (9a) depict the laser and its optics, whereas (8b) shows the LED panel for the backlight PIV. b) Reference for the height \(h\) of the measurement planes, measured from the container bottom.}
	 \label{fig:setup}
\end{figure}

Faraday waves are investigated in a circular container of acrylic glass (diameter 290 mm), similar to those used in pioneering studies ~\citep{von2011double, von2013measurement, francois2013inverse} filled with distilled water at  21\(\pm1\)	 \(^\circ\)C. A depth of 30 mm is chosen for a deep water approximation, such that the depth is larger than twice the wavelength of the Faraday waves at the fluid surface. The container is vertically shaken by an electromagnetic shaker (TIRA TV5220).
\noindent A schematic representation of the experimental set-up is \linebreak shown in Fig.~\ref{fig:setup}.

\noindent Monochromatic sine forcing at \(f_f= 50\)  Hz is imposed to the shaker from a function generator (RIGOL), which results in a Faraday wavelength \(\lambda_F\) of 9.5 \(\pm\) 1 mm and a subharmonic Faraday wave frequency response of 25 Hz. The acceleration of the container is measured with an accelerometer (Kistler, \(\pm5\) g, sensitivity 1 g/V\(\pm10\%\)) and read out by a high-frequency digitalizer (Spectrum). Two levels of forcing acceleration \(a_f\) are used: 0.47 and 0.70 g. The stronger forcing was chosen in order to achieve a highly homogeneous distribution of the surface wavefield in the whole container up to the boundary, while for the lower forcing some radial symmetry of the surface waves right at the boundary could be observed. The onset of Faraday waves is observed at a threshold acceleration of \(a_{\text{th}}=0.29\) g, and determined as described in \citep{ezersky1985chaotic} and \citep{tufillaro1989order} as a second threshold for the order-disorder transition from a highly-ordered and stable wave pattern to the formation of defects in such pattern. The threshold acceleration was averaged between the trheshold values observed by gradually increasing the forcing from standstill and those from reducing it from a stronger forcing. The resulting values for the dimensionless forcing amplitude,  defined as `supercriticality'  \(\varepsilon=(a_f-a_{\text{th}})/a_{\text{th}}\) \citep{francois2014three}, are \(\varepsilon=0.61\) and \(\varepsilon=1.41\) respectively.\\
In this study, measurements are carried out for a somewhat weaker forcing compared to the previous experiments (where a supercriticality of \(\varepsilon=1.7\) was chosen \citep{francois2014three}), since this was found to be an optimal setting with regard to wave height and image quality for particle image velocimetry. The upper bound of supercriticality in this experiments is the formation of water droplets forming on the surface of the waves.
\subsection{\label{sec:camera}Camera and Image Acquisition}
A second signal (standard TTL) from the function generator is used to trigger the high-speed camera (PCO dimax HS2, 12 bit depth). The camera is synchronised with the dominant frequency of the waves, which is found at the first subharmonic of the driving frequency \(f=f_f/2=25\) Hz, and records with a frame rate of 400 fps, corresponding to  $1/16 T$, $T$ beeing the wave period. The phase difference between the TTL driving the camera and the sine function fed into the shaker was then carefully monitored through the digitizer and tuned to capture the point of zero amplitude in the waves (flat surface) that occurs then at every eighth image. 
The camera is placed at the side of the shaker supports, and an optical prism-mirror is used to deflect the camera line of sight in the vertical direction. The camera resolution is 1400\(\times\)1050 pixels, and images had to be cropped to 1400\(\times\)850 pixels for the sub-surface measurements at \(a_f=0.70\) g to match the width of the laser sheet. Images are saved in a 16 bit  format (.b16), and subsequently converted back to a 12 bit format, which corresponds to the actual bit depth of the camera.  
\subsection{\label{sec:PIV_measurements}PIV Measurements}
\begin{figure}[h!]
		\centering
		\includegraphics[width=0.49\textwidth]{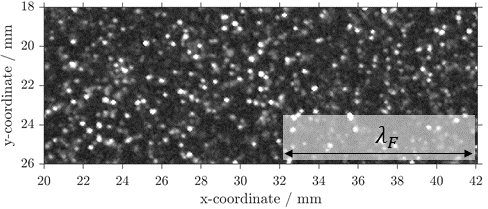}
		\caption{Example of raw images of the fluorescent particles from the case \(h=29\) mm,  \(a_f=0.70\) g.    A zoomed-in region of 350\(\times\)130 pixels is shown. The particles are 4-6 pixels in diameter and the Faraday wavelength \(\lambda_F\) is depicted for reference.}
		\label{fig:raw_data}
\end{figure}

Two PIV techniques are used for the measurements at and below the surface respectively, which mainly differ in the choice of illumination light source and tracer particles employed. For the measurements beneath the waver level, red fluorescent polyethylene microspheres are used (diameter of \linebreak 10-45 \(\mu\)m, Cospheric), illuminated by a continuous wave argon-ion laser (wavelength of 488 nm, Ion Technologies). An optical arrangement is used to deflect the laser beam (first upwards and later again horizontally) to the desired measurement height. Afterwards, the light passes through the light sheet optics (ILA 5150 GmbH) in order to generate a light sheet of 60 mm in width and 1 mm in thickness. 
The tracer particles have a density of 0.995 g/cm\(^3\) -  and uniformly disperse in the water volume, when additionally treated with a surfactant, as described below. The particles have a Stokes number \(St\sim\mathcal{O}(10^{-3})\), which indicates that the inertial effects are negligible and follow the flow well \citep{ouellette2008transport}. The particles are neutrally buoyant since after a day of disposal no accumulation of tracers at the bottom or at the surface is observed. A high-precision longpass filter (Edmund Optics) is used to capture the fluorescence of the particles (peak at 607 nm) and simultaneously shield the camera sensor from the laser light. The image of the  fluorescent particles on the camera chip results to be 2-6 pixels in diameter. The described settings were used to measure the velocity fields at multiple horizontal planes with height \(h\) from the container floor, \(h =\) 3, 4, 5, 10, 15, 20, 25, 26, 28 mm for the weaker forcing at \(a_f=0.47\) g, and \(h=\) 4, 10, 21, 27, 29 mm for the stronger forcing at \(a_f=0.70\) g, see Fig.~\ref{fig:setup} b). However, due to total light refraction at the water surface (\(h=30\) mm), the combination of laser and fluorescent particles could not be used to measure the velocity field right at the water surface itself. In this case, a combination of floating hollow glass microspheres (diameter of approx. 70 \(\mu\)m, \(St\sim\mathcal{O}(10^{-3})\), Fibre Glast) and a back-light (LED panel) was employed instead.\\
For both PIV techniques, 0.3 g of particles are wetted in a 10\%-solids solution with a surfactant (1\% Tween 80 solution, Polysorbate 80, non-ionic surfactant). This helps to uniformly disperse the naturally buoyant particles (fluorescent) in the water volume, and the same surfactant is used in all measurements in order to avoid differences in the waves (e.g. avoid changes in surface tension). With the available camera resolution (1400\(\times\)1050 px), the conversion factor for the spatial calibration of the field of view at the fluid surface \(h=30\) mm is 18.9436 px/mm for the case with \linebreak \(a_f=0.47\) g, resulting in a field of view of 73.90\(\times\)55.42 mm\(^2\). 
Fig.~\ref{fig:setup} provides a schematic representation of the experimental set-up for the two PIV techniques described above, and Fig.~\ref{fig:raw_data} shows an example of raw images of the fluorescent particles at \(h=29\) mm, \(a_f=0.70\) g. A zoomed-in region of 350\(\times\)130 pixels, corresponding to \(22\times8\) mm\(^2\) is shown and the Faraday wavelength \(\lambda_F\) is depicted for reference. 

\pagebreak

\section{Results and Discussion}
\label{sec:Results_Discussion} 
PIV measurements of the Faraday  flow are carried out on the  surface and at different depths in the water. Subsequently, the data is analysed using PIV View 2.6 with suitable interrogation window size and time intervals in order to have 4-6 particles per window and a particle displacement of 5-8 pixels. Due to the temporally well resolved measurements, these values could be adjusted by adapting the temporal step between successive images loaded to the PIV algorithm. The \(z\)-component of velocity  (normal to the light sheet planes) could not be reconstructed from the available set-up. 
For the following figures and diagrams, the notation \textbf{u}~\(=(u,v)^\top\) will be used to denote the velocity field and its components in \(x\)- and \(y\)-direction respectively, and \(h\) will be used for the height of the measurement plane with respect to the container bottom (see Fig.~\ref{fig:setup} b)). The background image in Fig.~\ref{fig:example_PIV} a) corresponds to the actual background-corrected images captured by means of backlight shadowgraphy, whereas in \ref{fig:example_PIV} b) the original background-corrected images have been inverted for better visualization of the velocity arrows.

\subsection{\label{sec:Vel_fields}Velocity Fields}
\begin{figure*}[b!]
	\centering
	    \begin{subfigure}[b]{0.49\textwidth}
        \centering
        		\includegraphics[height=0.67\textwidth ]{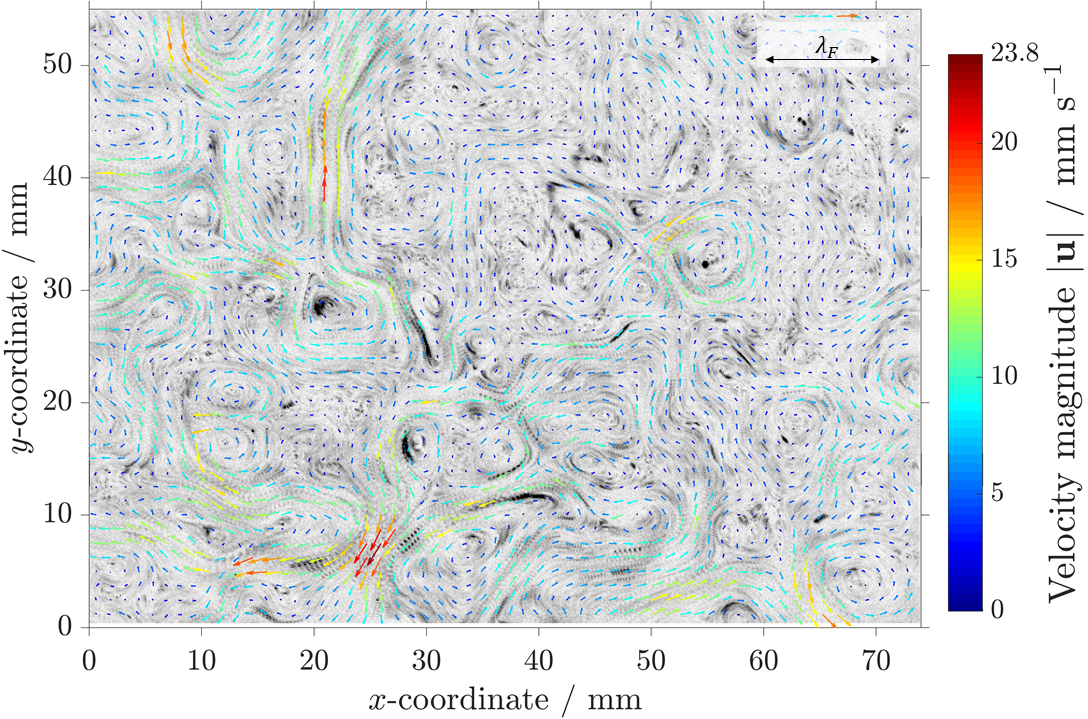}
        		\caption{\(h=30\) mm}
	\end{subfigure}
		\hfill
	    \begin{subfigure}[b]{0.49\textwidth}
        \centering
        		\includegraphics[height=0.67\textwidth ]{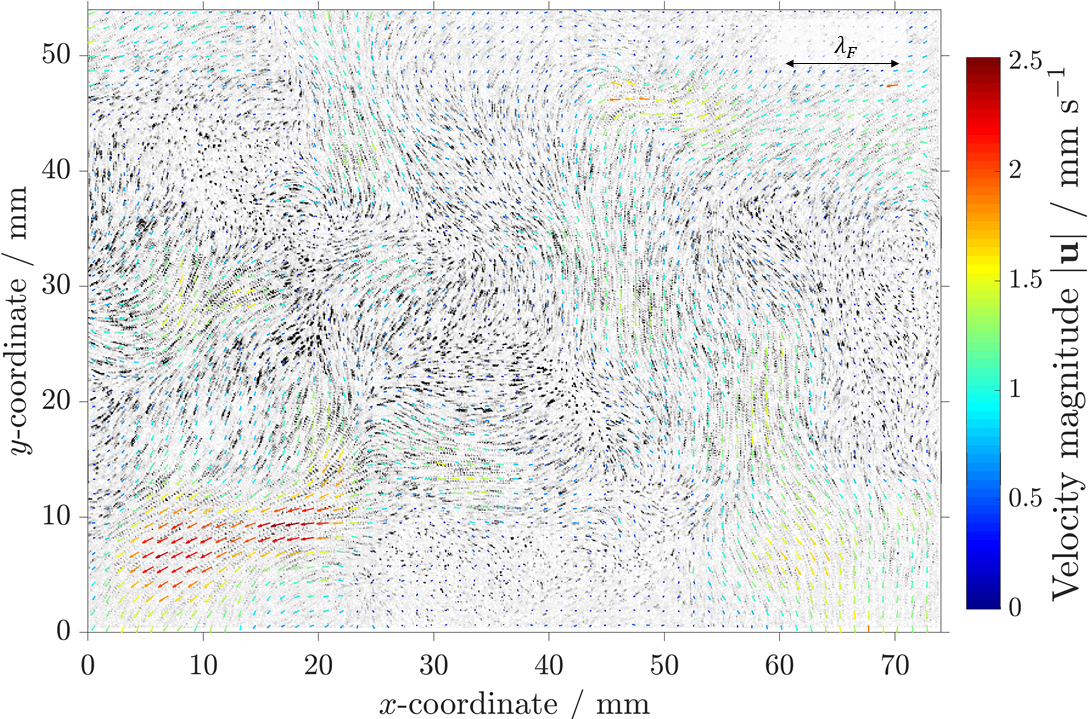}  
        		\caption{\(h=25\) mm}
	\end{subfigure}
	\caption{Representation of instantaneous velocity fields (every second arrow depicted) on top of the corresponding  background-corrected data averaged over 6 successive images for the \(a_f=0.70\) g case (color version online). a) Water surface, \(h=30 \) mm, conversion factor: 18.94 px/mm, velocity magnitude \(\vert \mbox{\textbf{u}}\vert\) ranging from 0 to 23.8 mm/s. b) Sub-surface, \(h=25\) mm, conversion factor = 18.96 px/mm, velocity magnitude \(\vert \mbox{\textbf{u}}\vert\) ranging from 0 to 2.5 mm/s, inverted background for better visibility of the colored velocity arrows. The Faraday wavelength \(\lambda_F\) is depicted for reference.}
	\label{fig:example_PIV}
\end{figure*}

Fig.~\ref{fig:example_PIV} shows an example of a velocity field for the \(a_f = 0.47\) g case at the water surface (\(h=30\) mm, Fig.~\ref{fig:example_PIV} a)), and 5 mm below the surface (\(h=25\) mm, Fig.~\ref{fig:example_PIV} b)). The background image is an average of 6 successive experimental frames and provides visual validation of the PIV calculations. \\

\begin{figure*}[!t]
	\centering
		\includegraphics[width=0.9\textwidth]{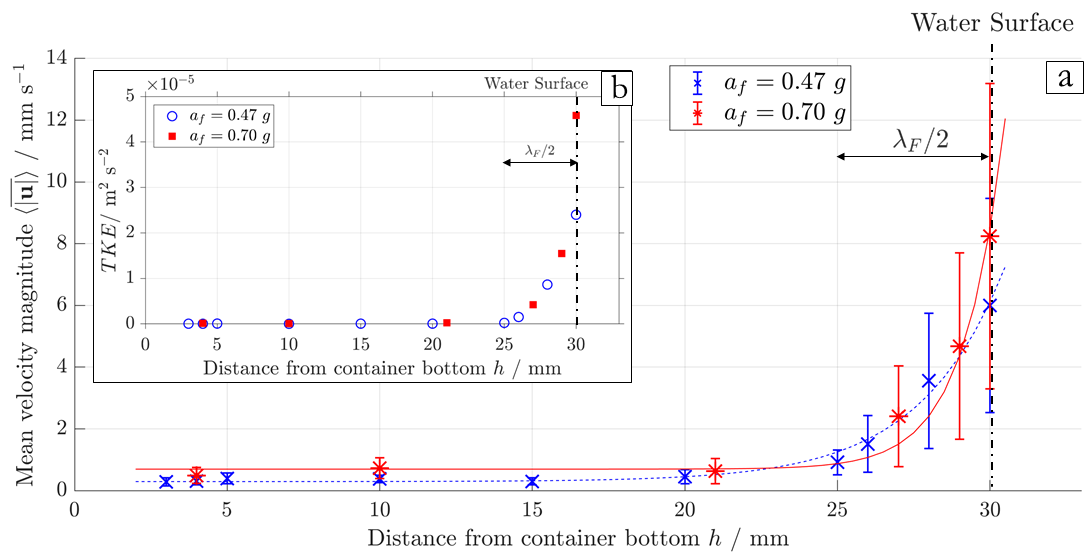}
	\caption{a) Profiles of mean velocity magnitude at different heights from container bottom for forcing amplitudes  \(a_f=0.47\) g 
	(blue \(\times\))  and \(a_f=0.70\) g (red *). Values averaged over all the time steps and grid points, with error bars in chart showing the standard deviation of the velocity magnitude. Exponential curves are fitted for both forcing cases (solid red line for  \(a_f=0.70\) g, blue dashed line for  \(a_f=0.47\) g) with a coefficient of determination \(R^2\geq 0.98\). 
	 Inset b): Profiles of \(TKE\) at different heights from container bottom for forcing amplitudes  \(a_f=0.47\) g (blue \(\circ\))  and \(a_f=0.70\) g (red-filled \(\square\)). Color version online, half Faraday wavelength depicted for reference, water surface level marked with dash-dotted line.}
	\label{fig:vel_profiles_vs_h}
\end{figure*}

\noindent From the velocity field, a few characteristics of the Faraday flow on the water surface can easily be recognised, namely the presence of multiple vortices with variable length scales, as also observed in \citep{von2011double}, as well regions of jet-like flow in which the flow is strongly accelerated, similar to the riverlike structures defined as ``trajectory bundles" in \citep{francois2018rectification}.
In contrast, it is evident from the depicted velocity-fields in Fig.~\ref{fig:example_PIV} b) that larger and slower structures exist below the surface (note the difference by an order of magnitude in the scale of velocity colorbars). The main difference from the flow field at the surface is that the vortices and the jet-like flow which are strongly related to the turbulent nature of the Faraday Flow are not observed for \(h<30\) mm. In a thin layer with a depth of \(\lambda_F/2\) below the surface, the structures gradually become less turbulent, and a dominant direction in the velocity field is seen in the available field of view throughout the entire measurement time.\\
Fig.~\ref{fig:vel_profiles_vs_h} a) depicts the profiles of mean velocity magnitude against the distance \(h\) from container bottom, where the values are space- and time-averaged (\(\langle . \rangle\) and \(\overline{(.)}\) respectively), although the total number of time steps and grid points vary with the height. At the surface, 624 time steps are available for the \(a_f=0.47\) g case (here presented in blue markers) and 1056 for \(a_f=0.70\) g (red markers). 
The inset graph b) shows the profile of turbulent kinetic energy (\(TKE\)), computed as \linebreak \(TKE=\frac{1}{2}\left(u'^2 + v'2\right)\), where \(u'\) and \(v'\) are the RMS values of the velocity fluctuations. 
A dramatic velocity reduction can be appreciated below the surface for both \(u\) and \(v\) components. In fact, the mean velocity magnitude drops to \(1/e\) of the surface value in 2-3 mm for both cases, and by a factor of about 6.5 in a thin layer of 5 mm right underneath the surface, which corresponds to half the Faraday wavelength.

\begin{figure*}[t!]
	\centering
	\includegraphics[width=0.9\textwidth]{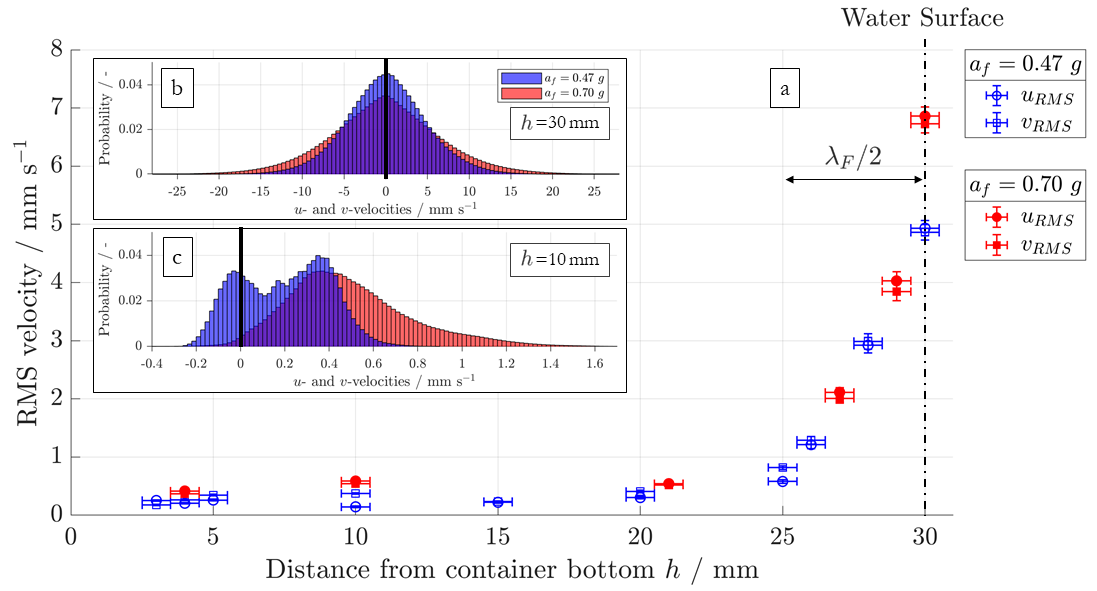}
	\caption{a) Profiles of RMS velocities at different heights. Values averaged over all the time steps and grid points, with error bars showing the error of measurement height \(\Delta h\) and velocity estimation derived from particle displacement  \(\Delta u_{\text{PIV}}\) for \(a_f=0.47\) g (blue, outline markers) and \(a_f=0.70\) g (red, filled markers), color version online. Inset b) and c): Combined probability distribution of \(u\)- and \(v\)-components of velocity for both forcing amplitudes and two different measurement heights (\(h=30\) mm and \(h=10\) mm respectively), averaged throughout the available timesteps and grid points. Forcing at  \(a_f=0.47\) g in blue,  \(a_f=0.70\) g in red. The 0-velocity is marked with a black solid line to highlight the symmetric distribution of velocities on the surface and the asymmetric velocity field below it. Color version online, half Faraday wavelength depicted for reference, water surface level marked with dash-dotted line.}
	\label{fig:vel_RMS_distr}
\end{figure*}

\noindent  For values of \(h\) below 20 mm the planar velocities level out at a nearly constant plateau. The  trend in the mean velocity magnitude profiles shown in Fig.~\ref{fig:vel_profiles_vs_h} a) could be well fitted with an exponential function. The goodness of the fit is confirmed by a coefficient of determination  \(R^2\geq 0.98\), which indicates the proportion of variation in the response variable \(\langle\overline{\mbox{{\textbf{|u|}}}}\rangle\) that can be explained by the independent variable (here \(h\)) in the linear regression model  \citep{devore2011probability}.
 The exponential decay of velocity magnitude is a somewhat unexpected result, since it corresponds to the most classical results of particles beneath travelling water waves \citep{dietrich1980general,breivik2014approximate} and the strong two-\linebreak dimensional turbulence at the surface seems to not alter this trend.\\
In Fig.~\ref{fig:vel_profiles_vs_h} a) the error bars indicate the standard deviation of the velocity magnitude signals, computed over all time steps and grid points, which gives an idea about how much larger the turbulent fluctuations are on the water surface compared to the sub-surface flow fields, which is confirmed by analysing the \(TKE\) profile, presented in Fig.~\ref{fig:vel_profiles_vs_h} b). Here, it is further shown that the flow structures developing below the turbulent surface layer barely show any turbulent kinetic energy below a depth of half the Faraday wavelength. Thus, the decrease in TKE is not merely caused by the decrease in mean flow velocity, but by the presence of large-scale and slower structures that persist longer in time that can be visually observed and are further analysed in Fig.~\ref{fig:autocorr_uu}.
Fig.~\ref{fig:vel_RMS_distr} depicts the profiles of the RMS velocities for both forcing amplitudes. These follow the trend of exponential decay as evaluated for the velocity magnitude. Furthermore, they reveal that the difference of RMS velocity values for \(u\) and \(v\)  increases below the surface. This was further investigated by looking at the probability distributions of the velocity components. The inset graphs in Fig.~\ref{fig:vel_RMS_distr} b) and c) show the combined probability distribution of \(u\) and \(v\) for both forcing amplitudes and at two different measurement heights (\(h=30\) mm and \(h=10\) mm respectively). The distributions are averaged through all available timesteps and grid points.
The error bars in Fig.~\ref{fig:vel_RMS_distr} a) indicate the range of  measurement uncertainties and error estimation. 
On the horizontal axis, an uncertainty in manually setting the measurement height is estimated at \(\Delta h=\pm 0.5\) mm. 
In the calculation of the velocities from PIV data, two sources of errors are considered, namely the calculation of particle displacement \(\Delta x_{\text{PIV}}\) and the determination of the spatial conversion factor from the calibration target \(\Delta x_{\text{cal}}\). The particle displacement error is estimated as explained in \citep{raffel2014particle}  and found to be \(\Delta x_{\text{PIV}}\leq0.1\) pixels.

\noindent This in turn yields a maximum uncertainty in the \(u\)- and  \(v\)-velocity estimation of \(\Delta u_{\text{PIV}}=0.16\) mm/s for the surface measurements at higher forcing (\(a_f=0.70\) g).
The error from the spatial calibration is estimated to be \(\Delta x_{\text{cal}}\pm2\) pixels over the total length of the calibration target, which ranges from 1114 to 1371 pixels. From this value a maximal uncertainty of \(\Delta u_{\text{cal}}=0.0011\) mm/s can be estimated on the surface, which is however negligible compared to the error from the particle displacement calculation and therefore only  \(\Delta u_{\text{PIV}}\) is shown.
In Fig.~\ref{fig:vel_RMS_distr} a) the RMS values of velocity are used to quantify   the distinct components since their mean value at the surface was found to be negligibly close to 0. Therefore, the velocity probability distributions presented in Fig.~\ref{fig:vel_RMS_distr} b) and c) are used to further characterize the visual observations described so far.
Here \(u\)- and \(v\)-components are combined together to improve the statistics).
The symmetric nature of the velocity fields on the surface can be appreciated, since for both forcing amplitudes the velocity probability  shows a symmetric distribution with respect to 0. The curve for the stronger forcing is also somewhat flatter since a broader range of velocities is reached and the TKE is higher. The situation is considerably different at \(h=10\) mm, where for the weaker forcing (blue bars) two distinct peaks can be recognized, which indicate different mean components of the velocity in \(u\) and \(v\), and a broad curve is seen for the stronger forcing with non-zero mean. For both forcing amplitudes the velocity probability distributions are asymmetric in planes below the surface, which indicates that the temporal persistence of the structures surpasses the experiment time and a much higher number of statistically independent realization of this experiment would have been necessary.

\subsection{Velocity Autocorrelation}
\begin{figure*}[t!]
	\centering
	\includegraphics[width=0.85\textwidth]{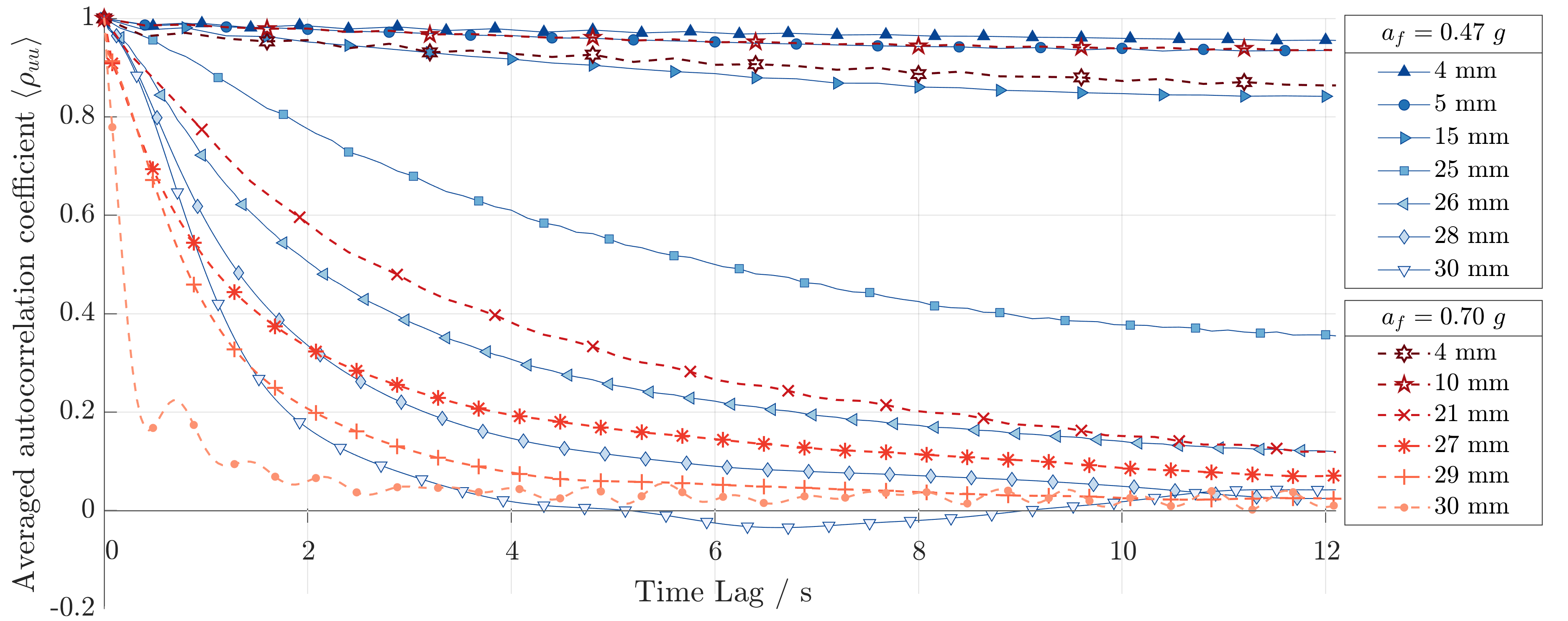}
	\caption{Spatially-averaged autocorrelation coefficient for the \(u-\) velocity fields at different heights \(h\) from the container bottom. Blue scale (solid lines) for \(a_f=0.47\) g and red scale (dashed lines) for \(a_f=0.70\) g (color version online). Both cases show a layer below the surface where the velocity correlations falls under the \(1/e\)-value in less than 2 s (and less than 1 s for the stronger forcing), indicating the presence of flow structures that decay and change over short time scales. However, velocity signals at \(h<20\) mm show a strong autocorrelation throughout the analysed time span and the horizontal velocities change only very slowly in direction and strength.}
	\label{fig:autocorr_uu}
\end{figure*}

In order to quantitatively analyse the long temporal persistence of the structures which was immediately observed by the eye, the autocorrelation of the velocity signals at different heights is analysed. The velocity autocorrelation is calculated in Fourier space  as presented in \citep{jahne2005digital} and converted back to space domain. Spatial averaging (denoted with \(\langle . \rangle\)) is carried out over all the grid points and every curve is then normalized with its maximum (at 0 time lag) to obtain the averaged autocorrelation coefficient \(\left\langle\rho\right\rangle\). 
The results are presented in Fig.~\ref{fig:autocorr_uu} for the \(u\)-velocity fields, namely \(\left\langle\rho_{uu}\right\rangle\), at different values of \(h\) for both forcing amplitudes.

\noindent  Very similar results have been obtained for the \(v\)-velocities (not shown here). In most cases at \(a_f=0.70\) g, depicted in red color scale and dashed line, data could be gathered over 2-3 replications of the experiments, which were statistically fully independent (the experiment has been restarted). Furthermore, the results are also averaged across the available cases. To further improve the statistics for the surface case the autocorrelation of \(u\) and \(v\) are averaged, as the velocity field is symmetric with respect to 0, as shown in Fig.~\ref{fig:vel_RMS_distr} b). From the autocorrelation diagram a trend can be immediately recognized, namely that the velocity structures change on short time scales in proximity of the surface, and that the decorrelation of velocity signals occurs much faster for stronger forcing, dropping below the \(1/e\)-value in under \linebreak 2 s for the \(a_f=0.47\) g case, and in under 1 s for the stronger forcing. The velocity fields however become extremely time-persistent close to the container bottom (lower values of \(h\)) for both forcing amplitudes.
 This trend is found to be consistent with the results from the velocity profiles from Fig.~\ref{fig:vel_profiles_vs_h} and \ref{fig:vel_RMS_distr}: at the water surface the flow is faster, with stronger turbulent fluctuations and structures that decorrelate more quickly in time. However, below a certain height \(h\) (25 mm for \(a_f=0.47\) g and 21 mm for \(a_f=0.70\) g) the strong autocorrelation indicates the presence of slower structures with time scales much longer than the time of the recordings.\\
For \(a_f=0.47\) g, the strong correlation below \(h=25\) mm was initially though to be related to border effects. In fact, for this forcing the distribution of the wavefield close to the borders at the surface also presents secondary flows in the radial direction, which in turn could lead to the development of different velocity layers below the surface. However, by increasing the forcing to \(a_f=0.70\) g, a homogeneously distributed wavefield is achieved on the whole surface, while the strong autocorrelation persist through the analysed time-span. This   confirms that the large-scaled structures below the surface develop independently from boundary effects.\\
It can also be noted that the autocorrelation at the surface for the stronger forcing presents an oscillating behaviour, which might be related to the stronger vertical motion of the waves. This behaviour is difficult to interpret by looking at velocity information on horizontal plane. \\
It is suspected that the horizontal motion of the particles on the whole surface, which is isolated by synchronising the camera frame rate with the vertical agitation, might not be entirely decoupled from the vertical shaking, and the oscillations in the autocorrelation thus result from slight 3D effects, in particular from a small residual convection in the \(z\) direction in the recorded images. This is also one of the main motivations for measuring at lower forcing amplitudes than \citep{francois2014three}  and it is reflected in the increasing values of divergence in the velocity fields on the fluid surface, considered in the next section.

\subsection{Vorticity and Divergence Fields}
\begin{figure*}[bh!]
	\centering
	    \begin{subfigure}[b]{0.48\textwidth}
        \centering
        		\includegraphics[width=\textwidth ]{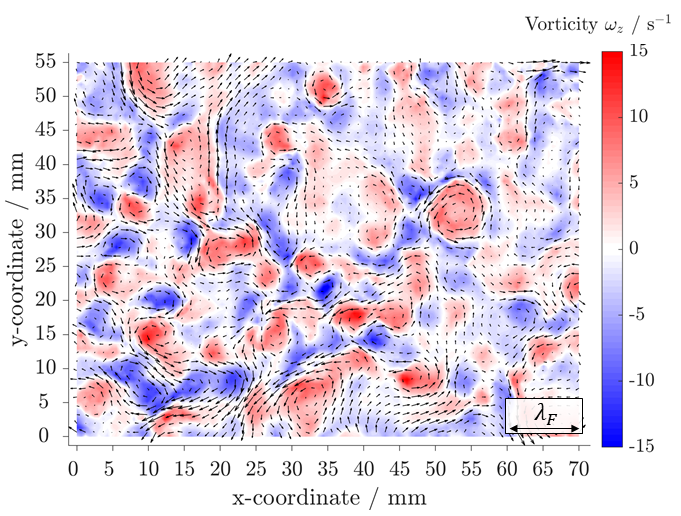}
        		\caption{\(\omega_z\) at \(h=30\) mm}
	\end{subfigure}
	\hfill
	\begin{subfigure}[b]{0.48\textwidth}
        \centering
        		\includegraphics[width=\textwidth ]{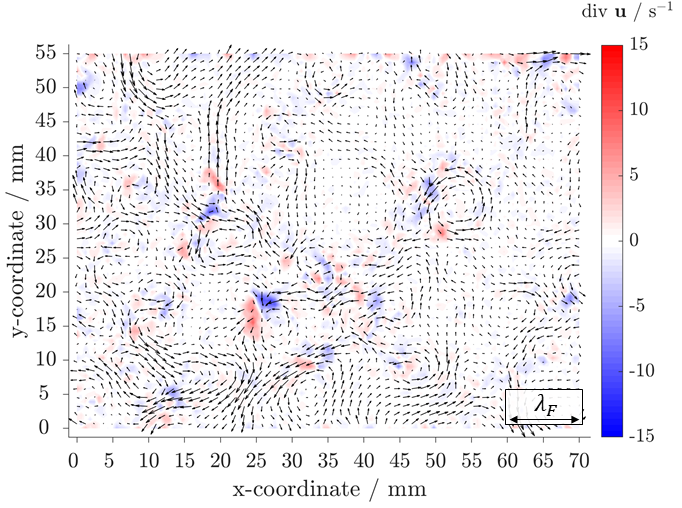}  
        		\caption{\(\text{div \textbf{u}}\) at \(h=30\) mm}
	\end{subfigure}
	\hspace{5pt}
		
	\begin{subfigure}[b]{0.48\textwidth}
        \centering
        		\includegraphics[width=\textwidth ]{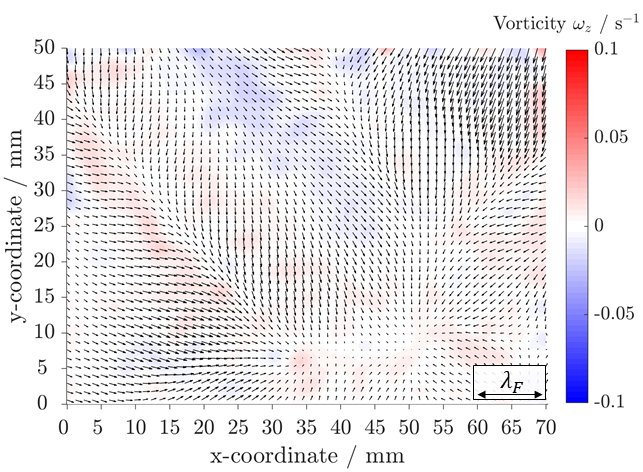}
        		\caption{\(\omega_z\) at \(h=4\) mm}
	\end{subfigure}
	\hfill
	    \begin{subfigure}[b]{0.48\textwidth}
        \centering
        		\includegraphics[width=\textwidth ]{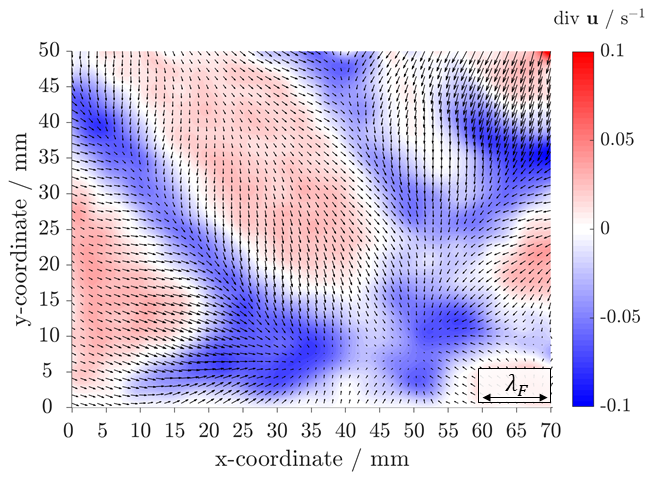}  
        		\caption{\(\text{div \textbf{u}}\) at \(h=4\) mm}
	\end{subfigure}

	\caption{Instantaneous vorticity (left) and divergence (right) fields at two different heights for a forcing \(a_f=0.47\) g. a), b) Surface measurements at \(h=30\) mm. c), d) Sub-surface measurements at \(h=4\) mm. The difference in spatial scales between values at the surface and below the surface can be appreciated. Black arrows qualitatively depict the local velocity field (vector length and colormap are scaled according to the height).  The Faraday wavelength \(\lambda_F\) is depicted for reference. Color version online.}
	\label{fig:div_vort_047g}
\end{figure*}

\begin{figure*}[th!]
	\centering
	    \begin{subfigure}[b]{0.49\textwidth}
        \centering
        		\includegraphics[width=\textwidth ]{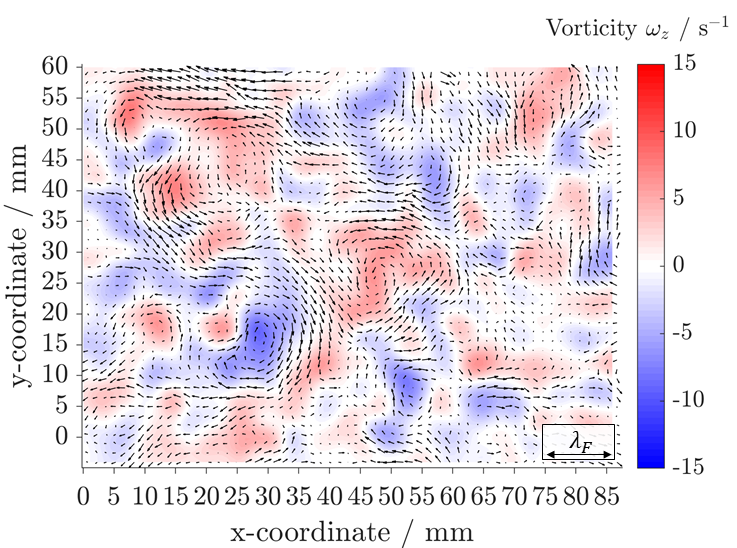}
        		\caption{\(\omega_z\) at \(h=30\) mm}
	\end{subfigure}
	\hfill
	    \begin{subfigure}[b]{0.49\textwidth}
        \centering
        		\includegraphics[width=\textwidth ]{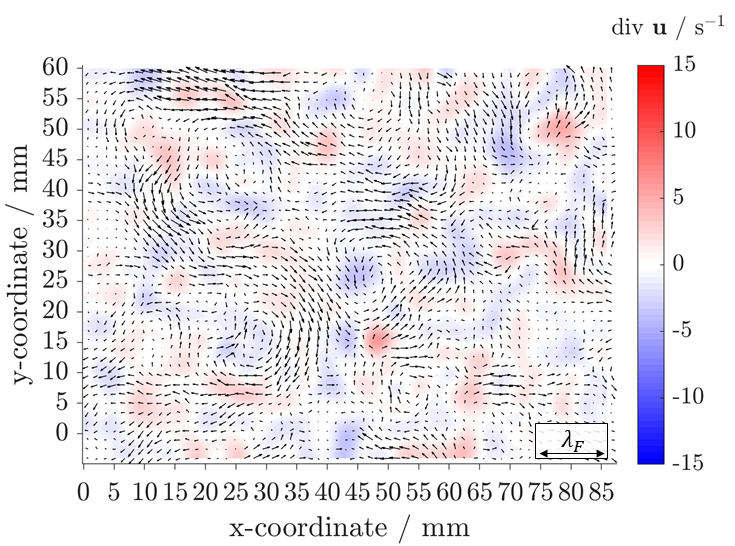}  
        		\caption{\(\text{div \textbf{u}}\) at \(h=30\) mm}
	\end{subfigure}

	    \begin{subfigure}[b]{0.49\textwidth}
        \centering
        		\includegraphics[width=\textwidth ]{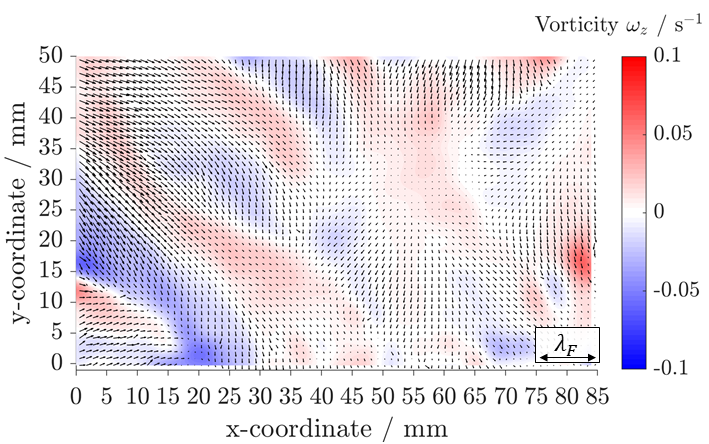}
        		\caption{\(\omega_z\) at \(h=4\) mm}
	\end{subfigure}
	\hfill
	    \begin{subfigure}[b]{0.49\textwidth}
        \centering
        		\includegraphics[width=\textwidth ]{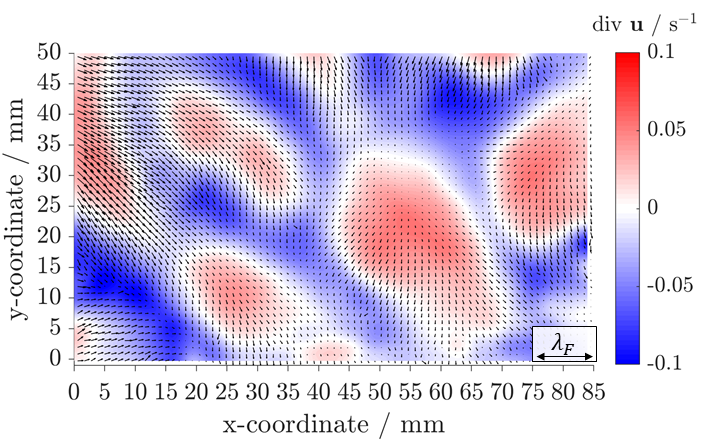}  
        		\caption{\(\text{div \textbf{u}}\) at \(h=4\) mm}
	\end{subfigure}
	\caption{Instantaneous vorticity (left) and divergence (right) fields at two different heights for a forcing \(a_f=0.70\) g. a), b) Surface measurements at \(h=30\) mm. c), d) Sub-surface measurements at \(h=4\) mm. The difference in spatial scales between values at the surface and below the surface can be appreciated. Black arrows qualitatively depict the local velocity field (vector length and colormap are scaled according to the height).  The Faraday wavelength \(\lambda_F\) is depicted for reference. Color version online.}
	\label{fig:div_vort_070g}
\end{figure*}

The relative size and behaviour of the ordered structures in the velocity fields is further investigated by analysing instantaneous contours of vorticity and divergence, computed for the 2D case as \(\omega_z= \partial v/\partial x-\partial u/\partial y\) and \(\text{div \textbf{u}}=\partial u/\partial x+\partial v/\partial y\) respectively.
The results are presented in Fig.~\ref{fig:div_vort_047g} for the \(a_f=0.47\) g case and Fig.~\ref{fig:div_vort_070g} for \(a_f=0.70\) g. The corresponding instantaneous velocity vectors are shown in black and the Faraday wavelength \(\lambda_F\) is included as reference. For the higher forcing, the velocity fields have been slightly smoothed with a Gaussian filter (with a kernel of \(3\times3\) velocity grid nodes) to remove measurement noise prior to calculating the sensitive gradients. The results reflect the findings from the velocity profiles and velocity autocorrelation. On the water surface, shown in  Fig. \ref{fig:div_vort_047g} a) and \ref{fig:div_vort_070g} a), regions of alternating vorticity peaks are densely distributed across the entire field of view, with a size varying mostly in between \(\lambda_F/2\) and \(2~\lambda_F\).
 Below the surface, the vorticity intensity drops significantly by more than one order of magnitude for both forcing amplitudes, as shown in Fig.~\ref{fig:div_vort_047g} c) and \ref{fig:div_vort_070g} c).  The vorticity structures become larger and smoother, and only a few, sparsely localized peaks can be recognized, which are not bounded by the structure of the Faraday waves.
 
 \noindent The divergence fields depicted in \ref{fig:div_vort_047g} and \ref{fig:div_vort_070g} b) and d) draw a different picture.  The divergence on the surface is overall lower than the vorticity, especially in the lower forcing case, further it presents only localized peaks and spatial scales much smaller than that of the vorticity structures.
Despite the non-zero value of divergence at the surface, it was shown in \citep{xia2017two} that the divergence (quantified by the compressibility parameter) rapidly decorrelates over four Faraday wave periods. This implies that the divergence results from the vertical movement of the waves  and does not in the mean contribute to the particle displacement, which in turn confirms  the quasi-two-dimensional nature of the Faraday flow. 
Nevertheless, the divergence fields for subsurface planes at \(h=4\) mm in Fig.~\ref{fig:div_vort_047g} and \ref{fig:div_vort_070g} d) indicate the presence of persistent vertical motion that in the horizontal-plane projection appears as a lattice of sinks and sources and resembles a pattern of convection rolls. The sinks are in this case narrow stripes of strong negative divergence, whereas the regions of positive divergence (flow from the bottom to the surface) are weaker but larger in size, and present an elliptical shape with a minor axis of approximately \(2\lambda_F\). This interpretation is supported  by the different patterns of streaming flows presented in \citep{gutierrez2016streaming} for purely longitudinal Faraday waves in shallow layers, whereas in our study the stripes additionally exhibit a temporal evolution, as their shape and orientation changes over time in a slow temporal scale.

\section{Conclusions and Outlook}
The Faraday experiment has been reproduced in a circular container with a diameter of 290 mm, filled up to a height of 30 mm with water. The container was vertically agitated with monochromatic forcing at a frequency of \(f_0=50\)  Hz and two forcing amplitudes  \(a_f=0.47\) g (\(\varepsilon=0.61\)) and \linebreak \(a_f=0.70\) g (\(\varepsilon=1.41\)). The resulting waves have a Faraday wavelength \(\lambda_F=9.5\pm1\) mm, and thus a deep water approximation for these waves is applicable (\(h>2\lambda_F\)). Time-resolved 2D-velocity fields have been measured with PIV techniques on the surface and at different horizontal planes in the water.\\
By analysing the profiles of mean velocity magnitude and velocity RMS in Fig.~\ref{fig:vel_profiles_vs_h} and \ref{fig:vel_RMS_distr}, it has been shown that for both forcing amplitudes the horizontal velocity decreases drastically beneath the water surface, such that at a depth of half the Faraday wavelength (\(h=25 mm\)) the mean absolute velocities are about 6.5 times smaller than on the surface). A velocity plateau is found for \(h<20\) mm. Astonishingly, the mean velocity magnitude profile is well fitted with an exponential function, as predicted from classical theory for the mean velocity below travelling water waves \citep{dietrich1980general,breivik2014approximate}. \\
The profile of turbulent kinetic energy presented in Fig.~\ref{fig:vel_profiles_vs_h} b) additionally reveals that the \(TKE\) sharply decreases below the surface, and that the turbulent velocity fluctuations are localized in a surface layer of thickness \(\lambda_F/2\). Additionally, the averaged velocity probability distributions depicted in Fig.~\ref{fig:vel_RMS_distr} b) and c) show that the flow on the surface is symmetric and smoothly distributed, whereas below the surface asymmetric distributions are observed, with distinct peaks at non-zero values that indicate dominant components of velocity in the observed region over the averaged time. This in turn shows the presence of slower and larger flow structures beneath the surface.\\
The analysis of velocity autocorrelation at different heights in Fig.~\ref{fig:autocorr_uu},  confirmed that the slower velocity structures below the surface also show  a longer persistence in time whereas the time scales of the turbulent flow on the surface are much shorter, as velocity signals decorrelate fast for both forcing amplitudes, falling well below the \(1/e\)-value in less than 2 seconds.\\
Finally, with the analysis of instantaneous vorticity and  divergence fields shown in Fig.~\ref{fig:div_vort_047g} and \ref{fig:div_vort_070g} it was possible to describe the instantaneous structures that develop beneath the surface and compare them to the ones that characterize the Faraday flow on the surface. The vorticity is considerably reduced and becomes small in comparison to the divergence. The vorticity structures become larger and smoother, and are not dominated by the structure of the Faraday waves as observed on the surface. Conversely, the divergence becomes stronger at the immersed planes, and persistent vertical motion can be seen in a  pattern of sinks and sources resembling convection rolls that change in shape and orientation in time on a slow temporal scale. The sinks appear as sharper ridge-lines with stronger negative divergence, which could also be visually observed during the experiments.  \\
To conclude, this article aims at clarifying the connection of Faraday flow and the vertical forcing due to the Faraday waves. Therefore we analyse subsurface velocity fields with respect to their two-dimensional motion and energy contents. Most importantly we find that the turbulent kinetic energy is confined to the surface and a shallow turbulent layer of thickness \(\lambda_F/2\) below it. This is in correspondence to other recent studies, that state that Faraday waves provide vertical oscillatory energy through a structure of oscillating solitons, denoted as the turbulent fuel of the surface flow \citep{francois2014three}. Furthermore, the existence of previously unknown slow and large scale structures beneath the surface is reported with unexpected temporal persistence and negligible levels of turbulence. A future study will consider the full three-dimensional velocity field to further unravel the interaction of Faraday waves, \linebreak 2D-turbulence at the surface and the persistent bulk flow.\\[6pt]
\newpage
\noindent \textbf{Funding}\\[3pt]
The authors gratefully acknowledge the financial support provided by the Deutsche
 Forschungsgemeinschaft (DFG) within the project 395843083 (KA 4854/1-1).\\[6pt]
\textbf{Acknowledgements}\\[3pt]  We thank Ms. Nikki Indresh Lal, M.Sc., and our colleague Mr. Sebastian  Hofmann, M.Sc., for the aid with the experimental set-up. \\[6pt]
\textbf{Conflict of interest}\\[3pt]
The authors declare that they have no conflict of interest.\\[6pt]
\textbf{Availability of data and material}\\[3pt]
Please contact the corresponding author to obtain access to the data.\\[6pt]
\textbf{Code availability}\\[3pt]
 Please contact the corresponding author to obtain access to the data.

%

\bibliographystyle{plainnat}
\bibliography{EIF_Subsurface_Measurements_of_Velocity.bib}   


\end{document}